\documentclass[12pt,a4paper]{article}

\usepackage[utf8]{inputenc}
\usepackage[T1]{fontenc}

\usepackage[a4paper,top=2cm,bottom=2cm,left=3cm,right=3cm,marginparwidth=1.75cm]{geometry}

\usepackage[affil-it]{authblk} 

\usepackage{amsmath}
\usepackage{graphicx}
\usepackage[colorlinks=true, allcolors=blue]{hyperref}
\usepackage{mathptmx}
\usepackage{titlesec}
\usepackage{siunitx}
\usepackage{booktabs}

\newcommand{\ra}[1]{\renewcommand{\arraystretch}{#1}}

\titleformat{\section}
  {\normalfont\fontsize{12}{15}\bfseries}{\thesection.}{1em}{}
\titleformat{\subsection}
  {\normalfont\fontsize{12}{15}\bfseries\itshape}{}{0.1em}{}

\titlespacing*{\section}
  {0pt}{1\baselineskip}{0\baselineskip}
\titlespacing*{\subsection}
  {0pt}{0\baselineskip}{0\baselineskip}

\usepackage{caption}

\usepackage{apacite}
\usepackage[english]{babel}

\usepackage{soul}
\usepackage[linesnumbered,ruled,vlined,lined,boxed,commentsnumbered]{algorithm2e}
\usepackage{amsmath}
\usepackage{amsthm}
\usepackage{mathtools}
\usepackage{algorithmic}
\usepackage{tabularx}
\usepackage[normalem]{ulem} 
\usepackage[textsize=small]{todonotes}

\usepackage{hyperref}
\hypersetup{
    colorlinks=true,
    linkcolor=blue,
    filecolor=magenta,      
    urlcolor=cyan,
}
\usepackage{multicol}
\usepackage[T1]{fontenc}     
\usepackage{amssymb}
\usepackage{subfig}

\newtheorem{defi}{Definition}

\newtheorem{assumption}{Assumption}

\addto{\captionsenglish}{%
    }
\title{\bfseries \normalsize Pooling for First and Last Mile: Integrating Carpooling and Transit}

\author[1]{Andrea Araldo*}
\author[2]{André dePalma}
\author[3]{Souhila Arib}
\author[1]{Vincent Gauthier}
\author[4]{Romain Sere}
\author[4]{Youssef Chaabouni }
\author[4]{Oussama Kharouaa}
\author[5]{Ado Adamou Abba Ari}

\affil[1]{Assoc. Prof., SAMOVAR, Télécom SudParis, Institut Polytechnique de Paris, France}
\affil[2]{Prof., CY Cergy Paris Université, France}
\affil[3]{Assoc. Prof., Computer Science Departement, CY Cergy Paris Université, France}
\affil[4]{AI Student, Computer Science Departement, CY Tech, France}

\date{\vspace{-5ex}}

\begin{document}
\maketitle

\begin{abstract}
While carpooling is widely adopted for long travels, it is by construction inefficient for daily commuting, where it is difficult to match drivers and riders, sharing similar origin, destination and time.

To overcome this limitation, we present an Integrated system, which integrates carpooling into transit, in the line of the philosophy of Mobility as a Service. Carpooling acts as feeder to transit and transit stations act as consolidation points, where trips of riders and drivers meet, increasing potential matching.

We present algorithms to construct multimodal rider trips (including transit and carpooling legs) and driver detours. Simulation shows that our Integrated system increases transit ridership and reduces auto-dependency, with respect to current practice, in which carpooling and transit are operated separately. Indeed, the Integrated system decreases the number of riders who are left with no feasible travel option and would thus be forced to use private cars. The simulation code is available as open source.
\end{abstract}


\section{Introduction}
In carpooling systems, a set of drivers accept to pickup and dropoff a set of drivers. Despite its success for inter-city trips, carpooling has not registered similar adoption for daily commuting in urban conurbations. Indeed, matching drivers and riders requires some ``sacrifice'' from them: they may both need to shift their departure and arrival times in order to ``meet'' at a time feasible for both; moreover, they have to change their routes, in order to meet at some meeting points. In daily commuting, the interurban time and route adjustments that users are willing to accept are much smaller than for long trips. These makes quite hard to match riders and drivers which have both similar, departure and arrival times and origins and destinations.

In this paper, we propose to overcome this limitations by adopting a Mobility as a Service philosophy. We show that Carpooling has limited benefit if managed independent from transit. Acknowledging the irreplaceable role of transit~\cite{basu2018automated}, we propose instead to integrate Carpooling into the transit offer. While integration of flexible modes into transit has been recently proposed~\cite{Calabro2021}, the integration of carpooling in particular has not been extensively studied. Few exceptions are~\cite{Stiglic2018} and~\cite{Fahnenschreiber2016}. However, the former assumes that riders obey to the matching proposed by the system, even if more convenient travel options were possible, which is unrealistic in practice. Moreover, they limit carpooling only in the First Mile (rider origin to transit station) and not in the Last Mile (station to rider destination). \cite{Fahnenschreiber2016}, instead, can only match one rider per driver.

We propose an Integrated System, which constructs via simple algorithms multimodal rider routes and driver journeys. We show in simulation \footnote{Code available at \url{https://github.com/YoussefChaabouni/Carpooling}} that such a system would provide a viable solution to private cars to a considerable number of commuters.

\section{System model}
\label{method}

\begin{figure}
    \centering
    \includegraphics[width=0.8\textwidth]{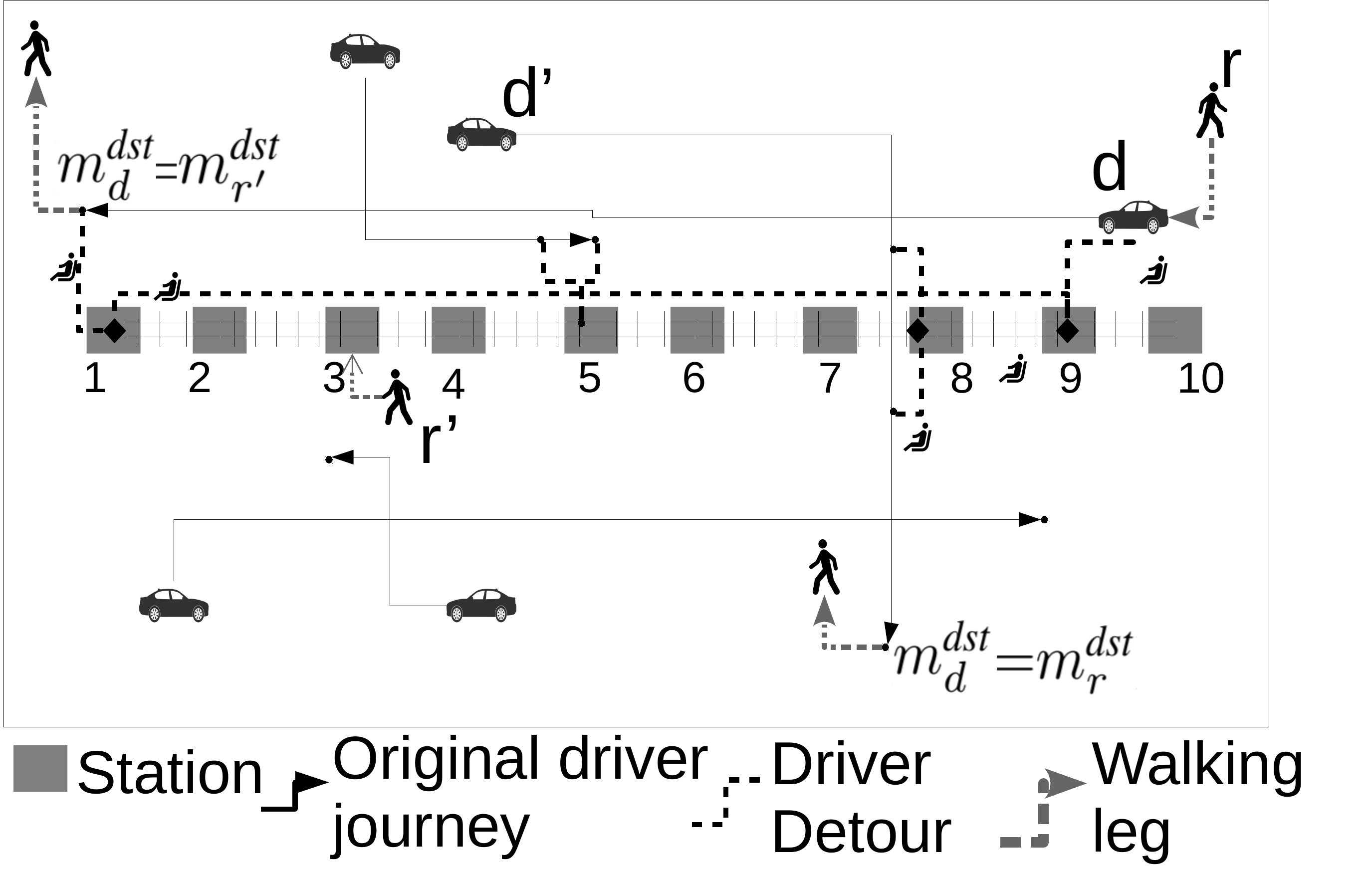}
    \caption{Illustrative scenario}
    \label{fig:scenario}
\end{figure}

We consider a suburban area, as in Fig.~\ref{fig:scenario}, served by a commuter rail. Users are either \emph{drivers} or \emph{riders}. Drivers are available to pick-up and drop-off other riders.

\label{sec:systems}
%
We compare three systems:
\begin{itemize}
    \item In the \textbf{\emph{No Carpooling System}} riders can just walk and/or use fixed schedule transit. 
    \item \textbf{\emph{Current System}}: like in current cities, carpooling and transit are handled separately, no multimodal trips are proposed by the system and driver journeys are completely independent from transit.
    \item We propose an \textbf{\emph{Integrated System}} in which the transit and carpooling routes are part of the same transportation service. Therefore, a rider can make a part of her trip by carpooling and the rest by transit. Driver journeys are integrated with transit via detours.
\end{itemize}


\subsection{Driver journey}
\label{sec:driver-journey}
The journey of driver $d$ is a sequence of \emph{meeting points} $m\in\mathcal{M}$. She can pickup or dropoff a passenger only in those. The origin and destination of driver $d$ are the first and last meeting points of her journey. All stations $\mathcal{S}$ are also meeting points, and thus a driver can potentially use them to start or to stop his/her journey.
%
%

In \textbf{No Carpooling} and \textbf{Current} System, driver $d$ just drives directly from her origin $m^\text{org}$ to her destination $m^\text{dst}$. In the Current system, she might pick passengers up in $m^\text{org}$ and drop them off at $m^\text{dst}$. In the \textbf{Integrated System}, a driver $d$ can make a detour to pass by $s^{org}_d$ or $s^{dst}_d$, i.e., the stations closest to origin $m^{org}_d$ and destination $m^{dst}_d$, or to pass by both. Such detours are accepted by driver $d$ only if her journey is no more than 15\% longer than the direct trip between $m^{org}_d$ and $m^{dst}_d$. The detour is realized only if there are riders to pickup or dropoff at the respective station, otherwise it is ignored. The calculation of the driver journey is detailed in Fig.~\ref{alg:driver-journey}.

We assume the incentive provided to the driver to make the proposed detours (possibly coming from riders' payments) is enough to accept. Incentive schemes~\cite{Zhong2020} are outside our scope.

\subsection{Transportation options available to riders}

\begin{figure*}[!h]
    \centering
    \includegraphics[width=1.05\textwidth]{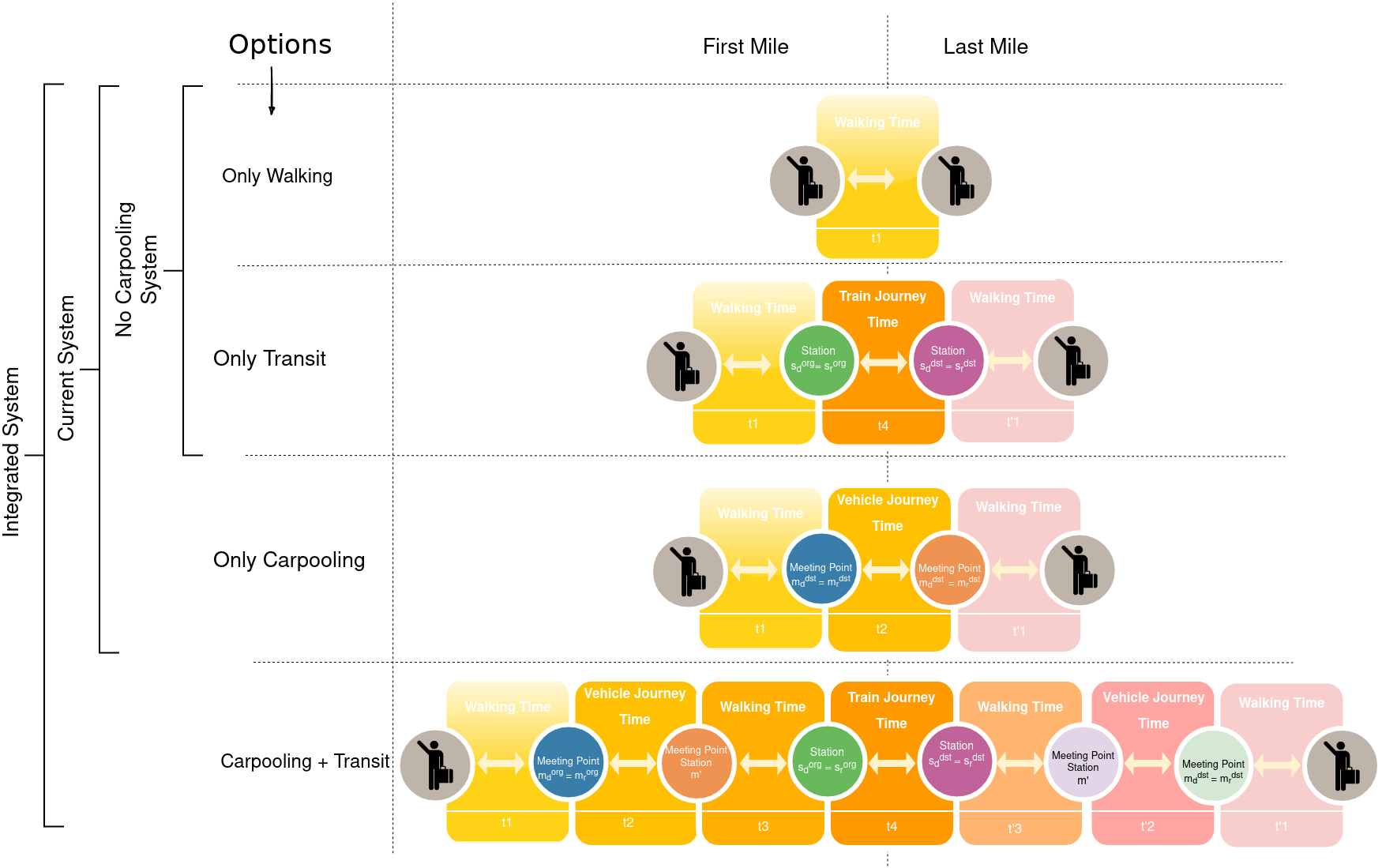}
    \caption{Options available for each of the three considered systems. }
    \label{fig00}
\end{figure*}

Let us assume journeys $J(d)$ of all drivers $d\in\mathcal{D}$ have been defined and consider rider $r$ departing at an origin $\textit{org}_r$ and willing to arrive to a destination $\text{dst}_r$ as soon as possible. The possible legs of a rider's journey, with the different options available in the three systems are depicted in Fig.~\ref{fig00}.

\begin{defi} 
\label{def:feasible}
A transportation option is \emph{feasible} for a rider only if (i) it implies a total waiting time of at most 45 minutes, (ii) a total walking distance of \SI{2.5}{\km} and (iii) if the total journey time is less than the one needed to go by foot from origin to destination.
\end{defi} 

In the previous definition, if a rider journey is composed by several legs, and she has to wait for several vehicles (either trains or drivers), the total waiting time is the sum of all waiting times. We assume that a rider aims to minimize waiting time by leaving home in order to arrive at a station or meeting point right at the moment where the vehicle she wants to board (carpooling or train) is departing.
Similarly, if the rider has multiple walking legs in a single journey, the total walking distance is the sum of their distance.

In the \textbf{No Carpooling System}, the rider has two possible \emph{transportation options}:
\begin{itemize}
    \item \emph{Only walking option}: Rider $r$ walks directly from $\textit{org}_r$ to $\text{dst}_r$.
    \item \emph{Only transit option}: Rider $r$ walks from her origin $\textit{org}_r$ to the closest station $s_r^\textit{org}\in \mathcal{S}$, waits for the next train, travels with that train to the station $s_r^\textit{dst}\in\mathcal{S}$ closest to her destination $\textit{dst}_r$, alights there and walks to $\textit{dst}_r$. 
\end{itemize}

In the \textbf{Current System}, in addition to the previous two, the following \emph{option} is available:
\begin{itemize}
        \item \emph{Only carpooling option}: Rider $r$ can carpool with a certain driver $d$ only if the origin and destination of $d$ are the closest meeting points to the origin and destination of $r$. In other words, if we define $m^\text{org}_r$,$m^\text{dst}_r$ as the meeting points closest to the origin and destination of rider $r$, we must have $m^\text{org}_r=m^\text{org}_d$ and $m^\text{dst}_r=m^\text{dst}_d$. If rider $r$ and driver $d$ carpool, $r$ first walks from her origin $\textit{org}_r$ to the origin meeting point $m_r^\textit{org}$, arriving right at the moment where $d$ is departing. Then, $r$ and $d$ carpool up to the destination meeting point $m_d^\text{dst}$ and, from there, $r$ walks to her final destination $\text{dst}_r$. Carpooling is possible if condition one holds (Def.~\ref{def:feasible}) and vehicle capacity is not exceeded.
        Among all the possible drivers with which rider $r$ can carpool, the system proposes the one that brings her to her final destination the earliest, via Fig.~\ref{alg:driver-selection}. 
\end{itemize}

In the \textbf{Integrated System}, in addition to the previous 3, the following \emph{option} is available:
\begin{itemize}
    \item \emph{Carpooling + Transit option}: Rider $r$ carpools (i) with a driver $d$ in the First Mile, i.e., from $r$'s origin $org_r$ to the closest station $s_r^{org}$, or (ii) with a driver $d'$ in the Last Mile, i.e., from the station $s_r^{dst}$ closest to $r$'s destination up to her destination $dst_r$, or (iii) with both $d$ in the First and $d'$ in the Last Mile.
    The system first computes the fastest way for rider $r$ to arrive to her closest station $s_r^org$. This can be either by only walking or by combining walking and carpooling. Then, rider $r$ takes the first train up to $s_r^{dst}$. The system finally computes the fastest way for rider $r$ to reach her final destination, which could be either by only walking or by carpooling with driver $d'$ and then walking. See the algorithm 4 describe in the figure~\ref{alg:carpooling+transit} for more details about the driver selection process used in our simulation.
    
\end{itemize}

Figure~\ref{fig00} summarizes the different transportation options available for riders. 
Observe that the trip depicted for Carpooling + Transit may also be shorter, in case the rider carpools only in the First or only in the Last Mile.

\begin{assumption} \label{as:1}

The system computes the earliest arrival time for each option and selects the one that allows the rider to arrive at her final destination the earliest.

\end{assumption}

\label{sec:selected}
If all the modes available to a rider are infeasible (in the sense of Def.~\ref{def:feasible}), then we consider her \textbf{unserved}: such users cannot use our system and need to resort to their private car.

Observe that the Integrated System offers more options to riders (cf. Fig.~\ref{fig00}). As a consequence, less riders will be unserved and the rider travel times decrease with respect to the No Carpooling and Current System. This will be confirmed by the numerical results.


\section{Performance Evaluation}
\label{sec:performance}

\subsection{Scenario Description}
The parameters of the scenario are in Table.~\ref{tab:scenario.table}. Observe that our users are not representative of the entire population of the area, but only of the ones that joined our system. The value of circuity (ratio between actual travelled distance from a point to another and euclidean distance) is taken from~\cite{Boeing2019}.

\begin{table}[!h]
    \ra{1.3}
    \centering
    \begin{tabular}{@{}llr@{}}
    \toprule
    Parameter & \phantom{abc} & Value \\
    \midrule
    \textbf{simulation area} && $15 \times$ \SI{8}{\km^2} \\
    \textbf{\# train station} && 10 \\
    \textbf{avg. distance between station} && \SI{1.5}{\km} \\
    \textbf{average speed} \\
    \hspace{5mm}walking && \SI{4.5}{\km\per\hour} \\
    \hspace{5mm}car (source: \url{statista.com}) && \SI{38}{\km\per\hour} \\
    \hspace{5mm}train && \SI{60}{\km\per\hour} \\
    \textbf{arrival density} \\
    \hspace{5mm}rider && \SI{8.3}{rider\per\km^2\per\hour } \\
    \hspace{5mm}driver && \SI{4.8}{driver\per\km^2\per\hour } \\
    \textbf{number of users} \\
    \hspace{5mm}riders && \SI{2988}{} \\
    \hspace{5mm}drivers && \SI{1728}{} \\
        \textbf{max. vehicle occupancy} && $4$ seats \\
    \textbf{network circuity} && 1.2\\
    \bottomrule
    \end{tabular}
    \caption{Simulation parameters}
    \label{tab:scenario.table}
\end{table}

We generated uniformly distributed meeting points $m_i\in\mathcal{M}$, with average density of \SI{3.55}{meeting\ point\per\km^2}. Then we added the train stations $s\in\mathcal{S}$ and also 4 to 5 meeting points uniformly distributed inside a circle of \SI{300}{\meter} radius centered  on each station $s\in\mathcal{S}$, to account for the higher population density therein. To generate origin $m^\text{org}_d$ and destination $m^\text{dst}_d$ of driver $d\in\mathcal{D}$, we select two random meeting points from the set $\mathcal{M}\setminus\mathcal{S}$.

We simulate driver and drivers departing in a 3h interval, but we only measure our metrics on the ones departing in the 1st hour, in order to avoid typical simulation boundary effects.



We contrast the No Carpooling and the Current systems with our Integrated System. To allow for direct comparison, we  provide the same input (i.e., the same set of rider origin-destination pairs and departure times and the same set of driver origin-destination pairs and departure times) to all the three systems.


\subsection{Served Transportation Demand}
\label{sec:served}

\begin{figure}[t]
    \centering
    \includegraphics[width=\textwidth]{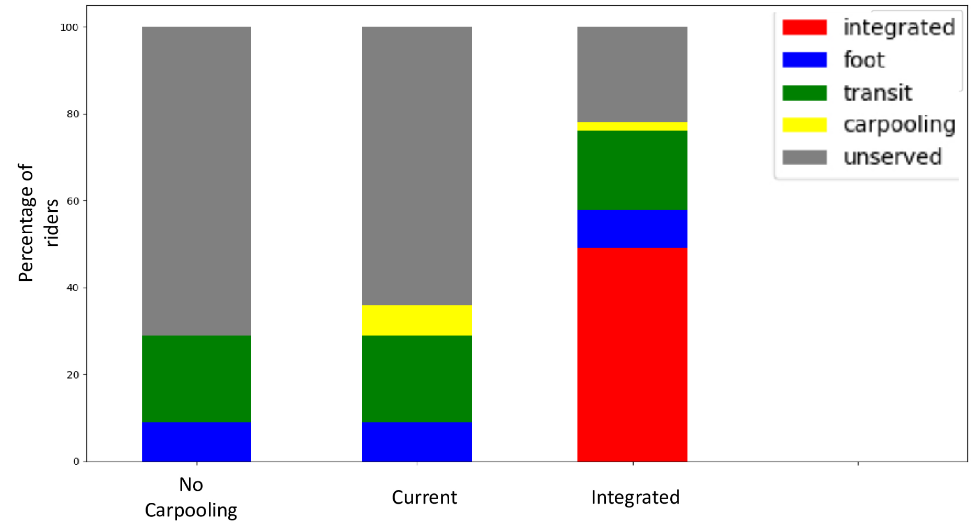}
    \caption{Breakdown of riders' transportation modes}
    \label{fig:passenger-breakdown}
\end{figure}

In Fig.~\ref{fig:passenger-breakdown}, we divide riders based on the selected multimodal transportation option. The percentage of unserved users in the Current System shows that carpooling itself is not beneficial without integrating it with transit. Indeed, only very few riders $r$ find a driver $d$ whose origin and destination meeting points correspond to hers ($m_d^\textit{org}=m_r^\textit{org}$ and $m_d^\textit{dst}=m_r^\textit{dst}$) and whose departure time is compatible with hers. We see instead that carpooling is an excellent feeder for transit: many riders find drivers to carpool with to reach a transit station in the First Mile or to go from a station to their destination in the Last Mile. In fact, transit stations take the role of demand consolidation points \cite{Araldo2019a}, which are easily served by carpooling.
A considerable amount of drivers, who were left with no feasible options in the Current system, thus being forced to take their private cars, can instead in the Integrated System perform their trip combining transit and carpooling.

\subsection{Travel Times}


\begin{figure*}[t]
\begin{center}
\subfloat[No Carpooling System]
{
    \includegraphics[width=0.33\textwidth]{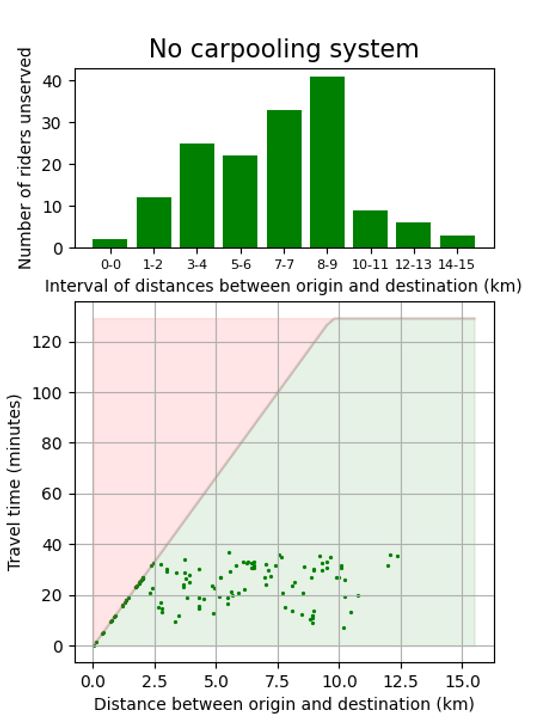}
    \label{fig:dist-tt-scatter-noc}
}
\subfloat[Current System]{
    \includegraphics[width=0.33\textwidth]{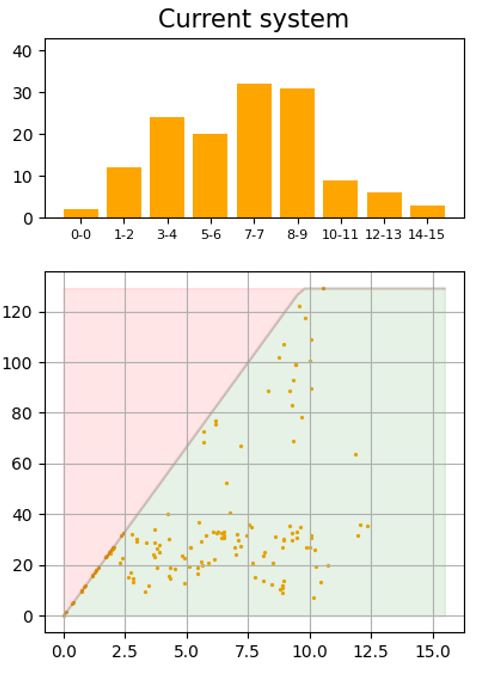}
    \label{fig:dist-tt-scatter-cur}
}
\subfloat[Our Integrated System]{
    \includegraphics[width=0.33\textwidth]{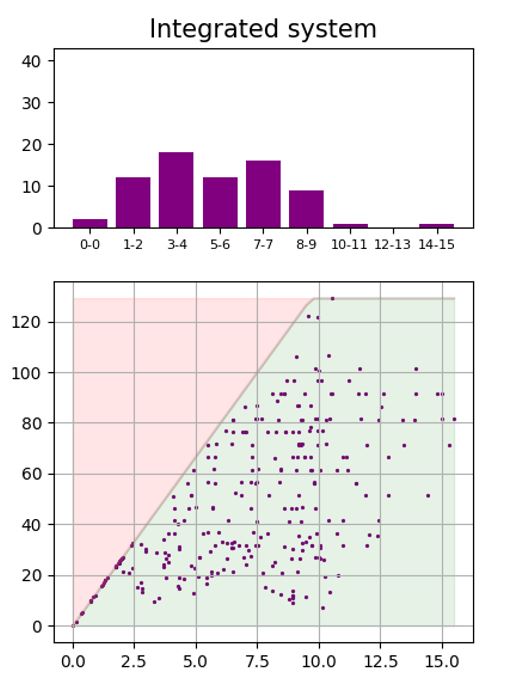}
    \label{fig:dist-tt-scatter-int}
}
\caption{Travel time against origin-destination distance. Each point represents a rider, the affine function is an average person walking. The upper bar plots represent the number of unserved riders, per each origin-destination distance value
}

\label{fig:tt}
\end{center}
\end{figure*}

In Fig.~\ref{fig:dist-tt-scatter-cur}, for each rider $r$, we plot her origin-destination distance against her travel time. As expected, almost all travelers manage to perform their trips for very short distances. However, as the distance increases, only a smaller part of them can do it.
In the No Carpooling System, no traveler can perform a trip of more than \SI{15}{\km}. Introducing carpooling, as in the Current System, creates feasible options for longer trips. However, only few ``lucky'' riders have such options, as the others do not find any driver with compatible origin, destination and departure time. The Integrated System, instead, provides feasible transportation options for much more travelers, which is particularly visible for longer trips. In general, for any distance, Fig.~\ref{fig:dist-tt-scatter-int} shows much more feasible trips in the Integrated System than in the other systems. The difference is given by the riders that were unserved in the other systems (upper bar plots) and have instead feasible options in the Integrated System.

\subsection{Driver vehicle Occupancy}

\begin{figure*}[h!]
\begin{center}
\subfloat[Current system]
{
    \includegraphics[width=0.9\textwidth]{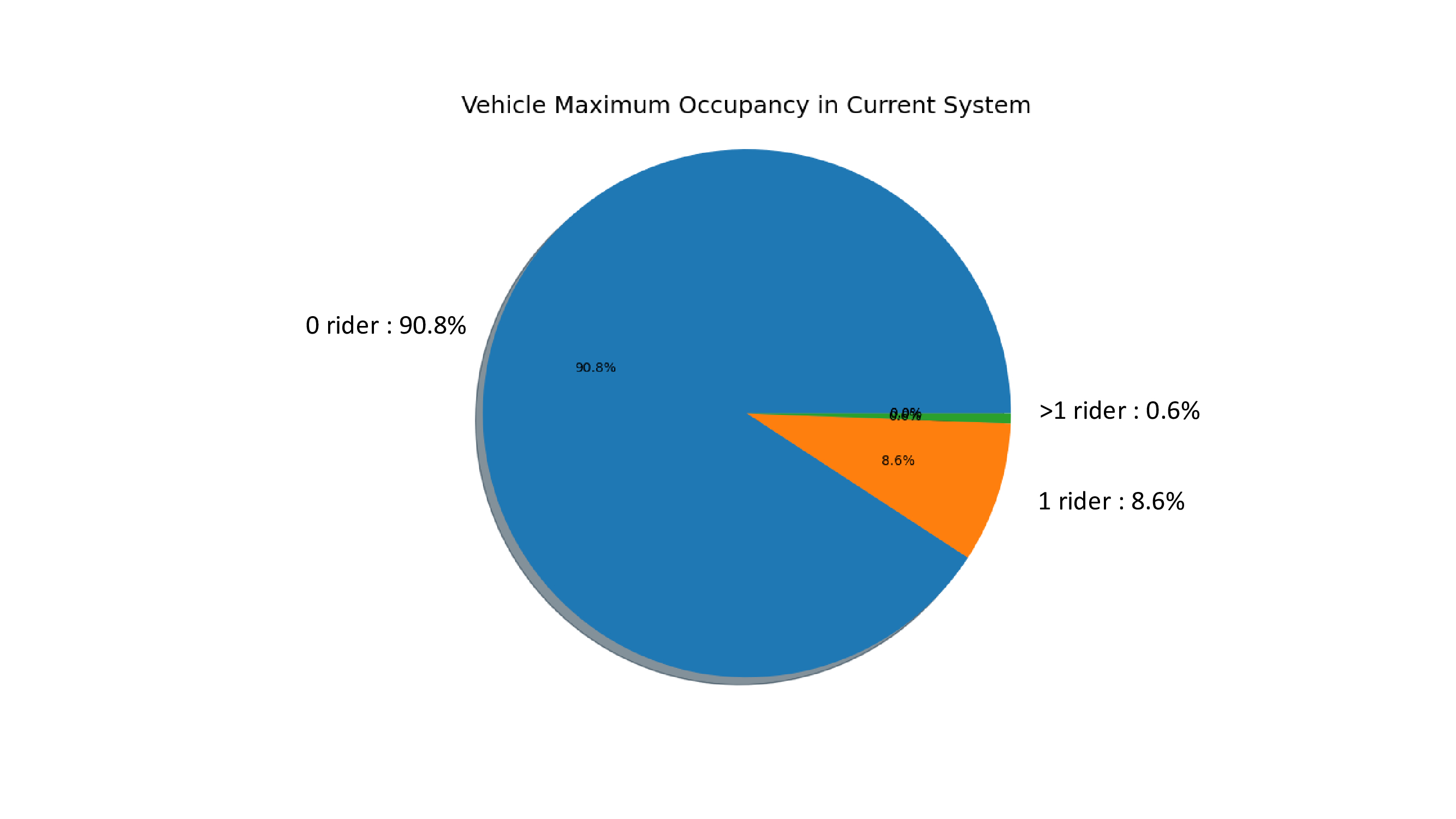}
    \label{fig:occupation-car-current}
}

\subfloat[Integrated system]
{
    \includegraphics[width=0.9\textwidth]{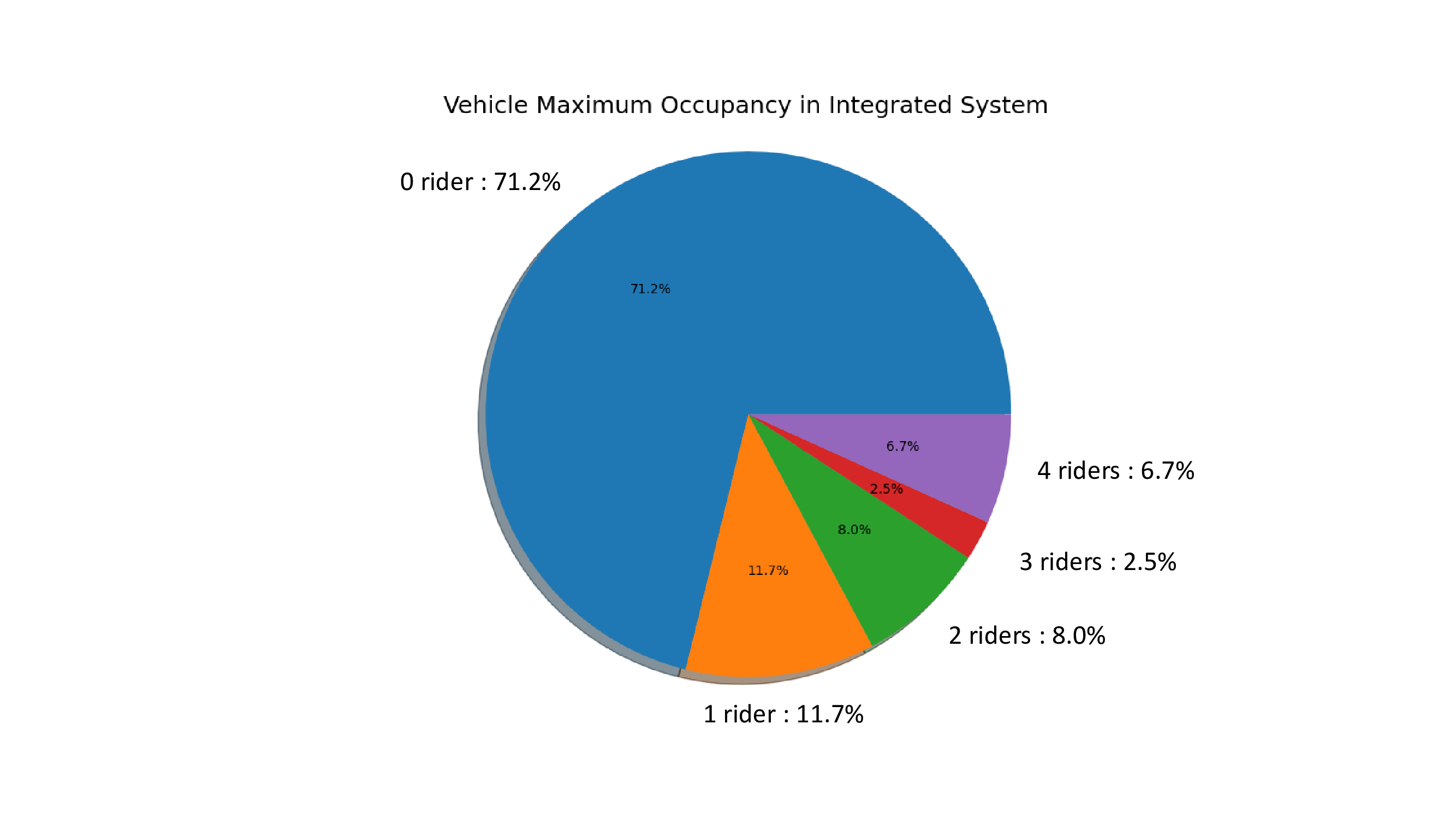}
    \label{fig:occupation-car-integrated}
}
\caption{Vehicle maximum occupancy in current and integrated systems}
\label{fig:occupation-car}
\end{center}
\end{figure*}

\begin{figure}[h!]
\begin{center}
{
    \includegraphics[width=0.9\textwidth]{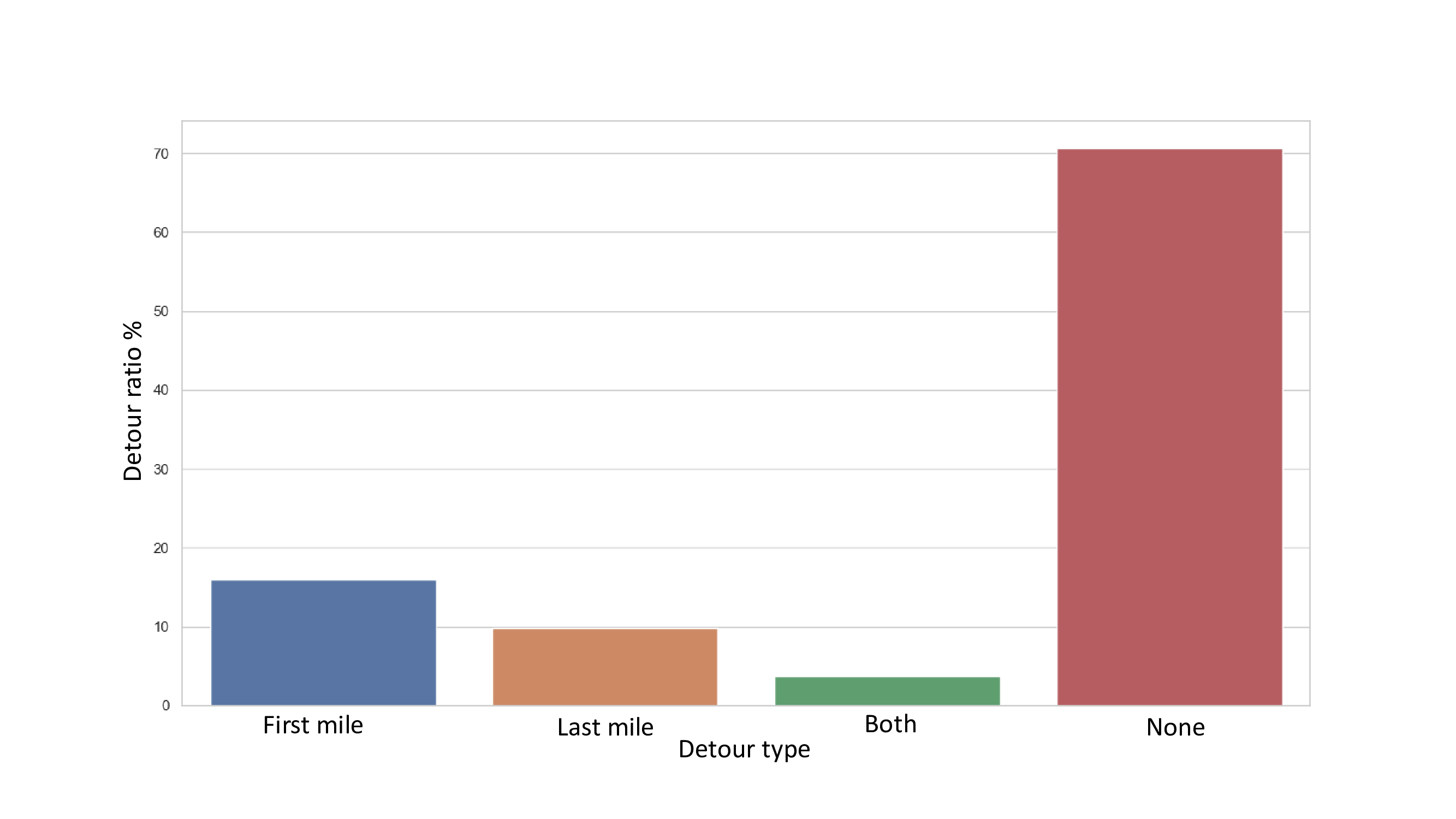}
    
}
\caption{Percentage of detours in the Integrated System}
\label{fig:detour}
\end{center}
\end{figure}

The number of occupied seats in a driver car changes with time, as riders board and alight. In figure~\ref{fig:occupation-car}, we focus on the \emph{maximum occupancy}, i.e., the maximum number of riders that have simultaneously been in driver $d$'s vehicle. 
%
The Integrated System allows to more efficiently exploit the capacity offered by carpooling. Indeed, in the Current System only very ``lucky'' riders $r$ find a feasible rider $d$ matching, i.e., with corresponding origin and destination meeting points ($m_d^{org}=m_r^{org}$ and $m_d^{dst}=m_r^{dst}$) and compatible departure times (neither too late nor too early). In the Integrated system, instead, transit stations are consolidation points, and the probability to find a driver passing by a station at the ``right'' time is relatively high. Observe also that this increase in rider-vehicle matching is also boosted by the fact that we purposely construct vehicle detours in order to preferentially pass through transit stations, around which we consolidate demand (riders) and offer (drivers), who can thus more easily be matched.

\subsection{Detours}
Fig. \ref{fig:detour} shows that the Integrated System requires detours only to a relatively small percentage of drivers, either in the first or last mile (i.e., through the station closest to the origin or the destination of the driver). Even fewer drivers make a detour in both first and last mile. This indicates that the Integrated System does not impose a high dis-utility to drivers. On the contrary, by just requiring relatively few driver detours, we are able to achieve high accessibility improvements for riders. This successful result is due to the demand consolidation operated around few meeting points and, more importantly, around transit stations.

\section{Conclusion}
\label{sec:conclusion}
We have proposed an Integrated System in which carpooling and transit are offered as a unified mobility service. 
By requiring relatively small detours to drivers, our system greatly increases  accessibility and richer feasible travel options, which would allow to reduce the need for using private cars, with societal and environmental benefits.


\section*{Acknowledgement}
Supported by the Paris Ile-de-France Region (DIM RFSI). 

\bibliographystyle{apacite}
\bibliography{sample}

\section{Appendix}
\label{sec:algorithms}

\begin{figure*}[!h]
\begin{center}
{
    \includegraphics[width=1\textwidth]{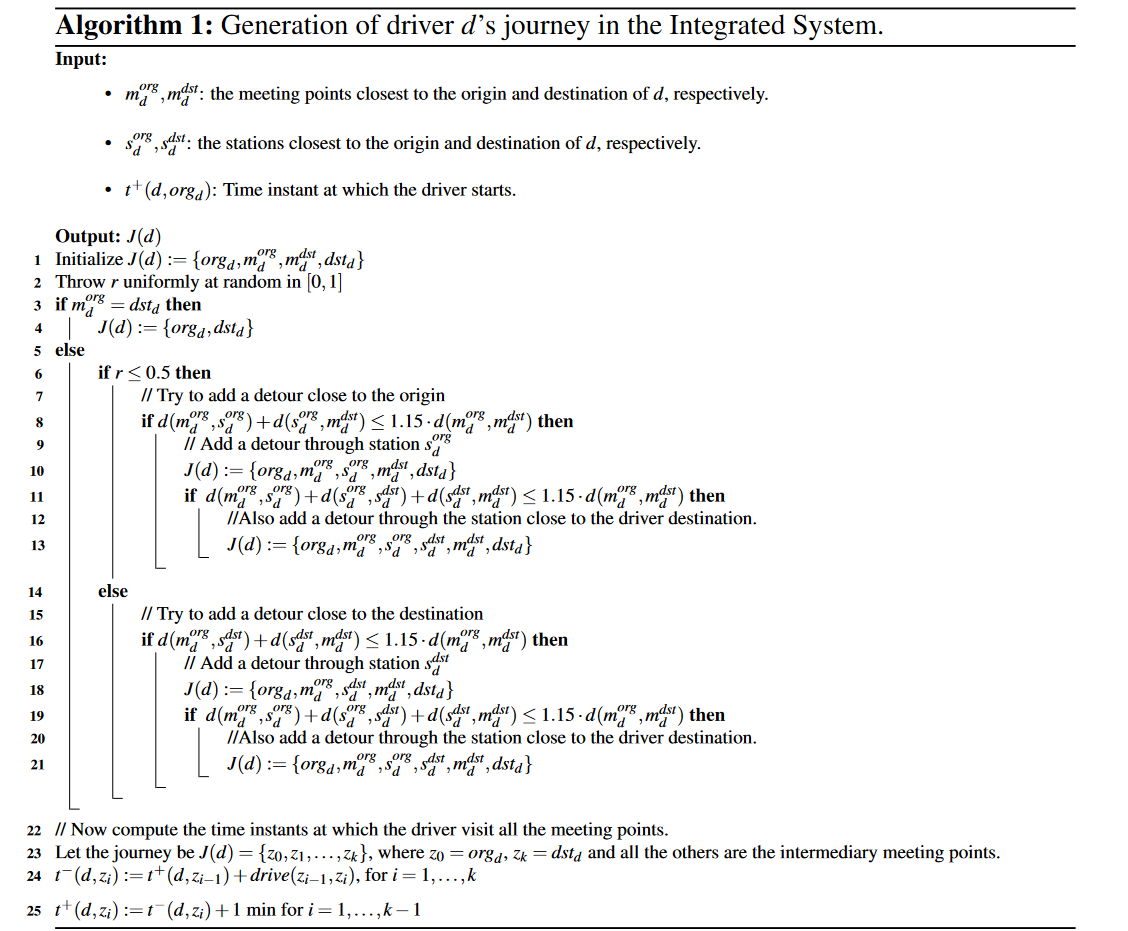}
    
}
\caption{Pseudo-code for the algorithm 1}
\label{alg:driver-journey}
\end{center}
\end{figure*}

\begin{figure*}[!h]
\begin{center}
{
    \includegraphics[width=1\textwidth]{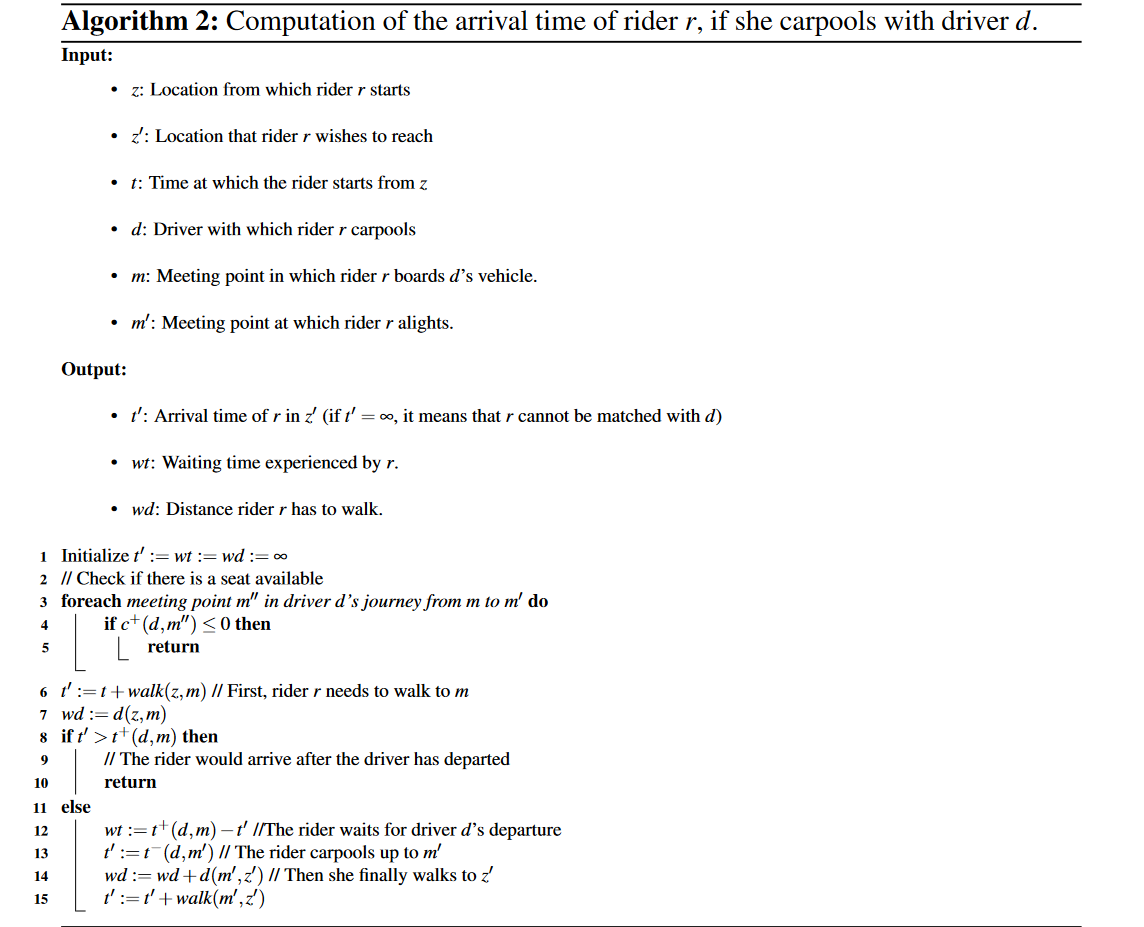}
    
}
\caption{Pseudo-code for the algorithm 2}
\label{alg:arrival-time}
\end{center}
\end{figure*}

\begin{figure*}[!h]
\begin{center}
{
    \includegraphics[width=1\textwidth]{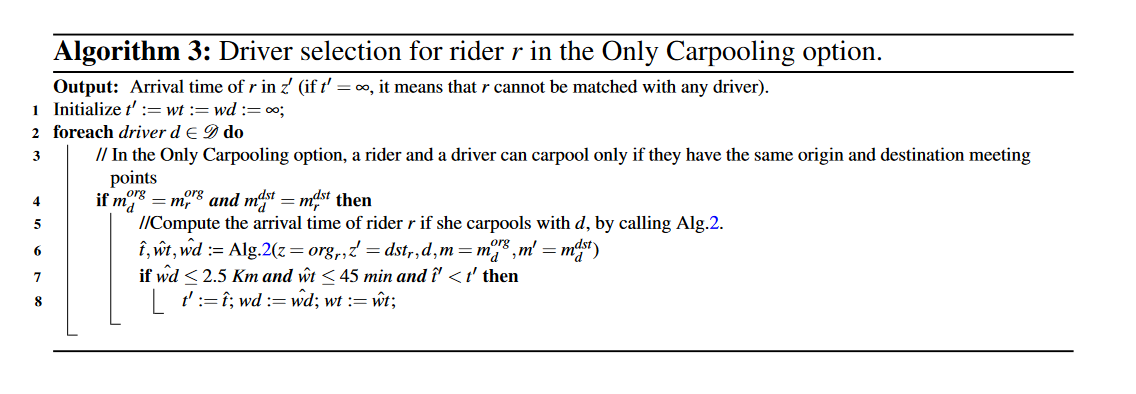}
    
}
\caption{Pseudo-code for the algorithm 3}
\label{alg:driver-selection}
\end{center}
\end{figure*}

\begin{figure*}
\begin{center}
{
    \includegraphics[width=1\textwidth]{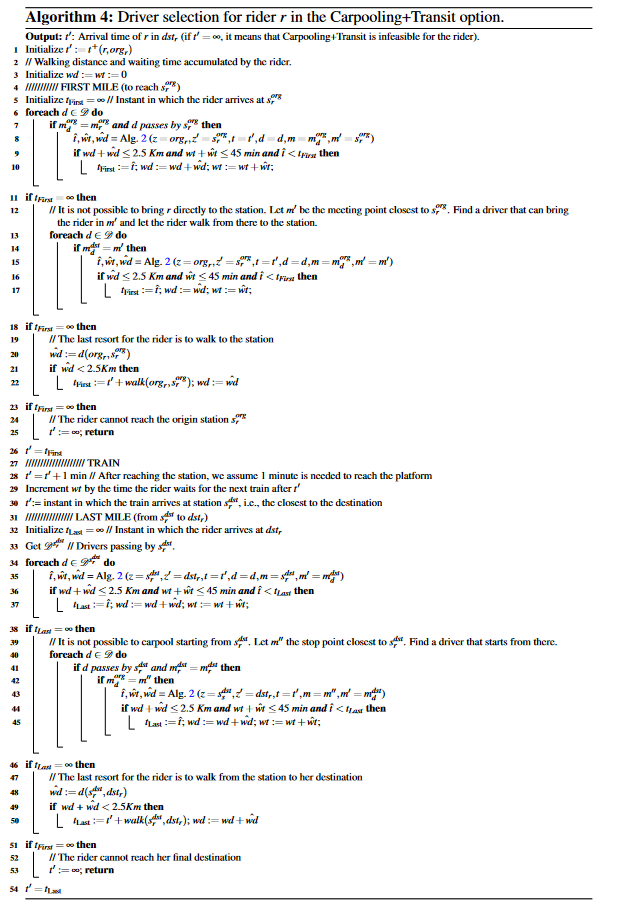}
    
}
\caption{Pseudo-code for the algorithm 4}
\label{alg:carpooling+transit}
\end{center}
\end{figure*}

We report now Fig.~\ref{alg:driver-journey}, \ref{alg:arrival-time}, \ref{alg:driver-selection} and \ref{alg:carpooling+transit} run by the Controller.

\end{document}